\documentclass[preprint,12pt]{elsarticle}

\usepackage{amssymb}
\usepackage{natbib,hyperref}
\usepackage{rotating}  
\usepackage{adjustbox}  
\usepackage{tabularx}
\usepackage{caption}  
\usepackage{booktabs} 
\usepackage{makecell}

\journal{tbd}

\begin{document}

\begin{frontmatter}



\title{Efficient speech detection in environmental audio using acoustic recognition and knowledge distillation}


\author[inst1]{Drew Priebe}

\affiliation[inst1]{organization={Department of Cognitive Science and Artificial Intelligence, Tilburg University},
            city={Tilburg},
            country={The Netherlands}
            }

\author[inst2]{Burooj Ghani}
\author[inst1,inst2]{Dan Stowell}

\affiliation[inst2]{organization={Naturalis Biodiversity Center},
            city={Leiden},
            country={The Netherlands}
            }

\begin{abstract}
The ongoing biodiversity crisis, driven by factors such as land-use change and global warming, emphasizes the need for effective ecological monitoring methods. Acoustic monitoring of biodiversity has emerged as an important monitoring tool. Detecting human voices in soundscape monitoring projects is useful both for analysing human disturbance and for privacy filtering. Despite significant strides in deep learning in recent years, the deployment of large neural networks on compact devices poses challenges due to memory and latency constraints. Our approach focuses on leveraging knowledge distillation techniques to design efficient, lightweight student models for speech detection in bioacoustics. In particular, we employed the MobileNetV3-Small-Pi model to create compact yet effective student architectures to compare against the larger EcoVAD teacher model, a well-regarded voice detection architecture in eco-acoustic monitoring. The comparative analysis included examining various configurations of the MobileNetV3-Small-Pi derived student models to identify optimal performance. Additionally, a thorough evaluation of different distillation techniques was conducted to ascertain the most effective method for model selection. Our findings revealed that the distilled models exhibited comparable performance to the EcoVAD teacher model, indicating a promising approach to overcoming computational barriers for real-time ecological monitoring.

\end{abstract}




\end{frontmatter}


\section{Introduction}
\label{sec:sample1}

Bioacoustics is the scientific discipline that focuses on sounds generated by animals~\citep{Stowell2022computational}. The field offers insight into the behaviours, communication, and migration patterns of different species. Recent advances in computational bioacoustics, such as data storage, and digital recording costs, enable the application of more advanced analytical approaches like deep learning~\citep{Stowell2022computational}. While early deep learning methods focused on neural networks such as the multi-layer perceptron, Convolutional Neural Networks (CNN) and Recurrent Neural Network (RNN) models currently surpass and exceed MLP models in the field~\citep{Stowell2022computational}. More recently, a convolution-free Audio Spectrogram Transformer (AST), an attention-based model for audio classification was designed~\citep{gong2021ast}. However, due to the quadratic complexity of self-attention, transformer-based models such as AST are known to be computationally expensive, resulting in increased latency and model size when compared to lightweight CNNs~\citep{pan2022edgevits}.

Despite the recent progress in computational bioacoustics, some practical and theoretical obstacles remain that prevent deep learning methods from broad usage in the field. A notable obstacle arises from the intricacies of dealing with human speech recordings in wildlife settings. Although these recordings serve as a useful proxy for quantifying human disturbance on ecosystems, they also allow for a more precise assessment of human presence~\citep{cretois2022voice}. This increased precision, while beneficial in one aspect, could lead to significant data privacy concerns as acoustic monitoring equipment becomes more advanced and more widely implemented~\citep{Stowell2022computational}. The implications of this obstacle extend even further, given the documented impact of human activity on the temporal dynamics of animal activity patterns, which include an increase in nocturnality and potential consequences for ecological interactions~\citep{gaynor2018influence, lewis2021human, hoke2021spatio}.
Moreover, noise pollution levels in protected areas have doubled, affecting critical habitat areas for endangered species~\citep{buxton2017noise}. In response to these challenges, Cretois, Rosten, and Sethi~\citep{cretois2022voice} developed a voice activity detection (VAD) model, EcoVAD, aimed at addressing both the need for precise measurement of human presence and privacy preservation in eco-acoustic data.

In addition to the above mentioned theoretical challenges, there are practical challenges that prevent deep models, such as EcoVAD, from being deployed in eco-acoustic environments. The deployment of such models have high latency costs~\citep{Stowell2022computational}. The current state-of-the-art acoustic monitoring tool, AudioMoth~\citep{hill2019audiomoth}, is a low cost, low power solution to certain technical challenges in bioacoustics. However, AudioMoth is not efficient enough to execute Deep neural networks (DNNs) in real time~\citep{solomes2020efficient}. The challenges that DNNs bring for deploying models on small devices has lead to a series of model compression and acceleration techniques, one of which is knowledge distillation~\citep{gou2021knowledge}. The main idea behind knowledge distillation is that a student model is trained to emulate the processing done by a larger teacher model, in order to distill refined knowledge and obtain a competitive performance versus the teacher~\citep{gou2021knowledge}. This technique allows efficient DNNs to be trained from large DNNs without a substantial drop in accuracy~\citep{gou2021knowledge}. 

While knowledge distillation addresses the compression framework required for deployment on edge devices, architectural efficiency remains another critical aspect for real time inference~\citep{hershey2017cnn, Stowell2022computational}. In an attempt to design a more efficient architecture, Howard et al.~\citep{howard2017mobilenets} introduced the MobileNetV1 architecture, which replaced the convolutional layer of CNN with depth wise separable convolutions. Specifically, the utilisation of factorised convolutions through the combination of depth-wise and point wise convolution, reduced the computation required by the convolutional block by a factor of 8~\citep{howard2017mobilenets}. While the introduction of MobileNetV1 allowed for a reduction in parameters without a significant loss in accuracy, it was not effective at efficiently extracting the manifold of interest (MOI)~\citep{sandler2018mobilenetv2}. This issue was in part due to the application of the non-linear functions (RELU) on low dimensional activations, which lead to information loss in the MOI. To confront this problem within the MobileNetV1 architecture, ~\citep{sandler2018mobilenetv2} introduced MobileNetV2, which incorporated inverted residuals with a linear bottleneck. In order to improve the representational power of the CNN architecture, Hu, Shen, and Sun~\citep{hu2018squeeze} implemented a Squeeze-and-Excitation block (SE), which allows the weighting of inter-dependencies between channels for feature selection. In light of this development, researchers then attempted to augment MobileNetV2, and introduced the SE block in the MobileNetV3 architecture. As a result, this led an improvement in both latency and parameter size of the model~\citep{howard2019searching}. 

Despite the design of MobileNet architectures addressing the model complexity and latency costs for deployment on small mobile devices, these architectures are not optimised for other edge devices, such as Raspberry Pi, NVIDIA Jetson Nano, and Google Coral, which contain different hardware specifications~\citep{kang2022evaluation, glegola2021mobilenet}. In an attempt to improve the MobileNetV3 design for Raspberry Pi devices, MobileNetV3-Small-Pi was developed~\citep{glegola2021mobilenet}. This architecture replaced the 5x5 filter with a 3x3 filter in the convolution block, and changed the hard-swish activation function to RELU. The modifications made to the MobileNetV3 lead to improvement in both latency and accuracy in MobileNetV3-Small-Pi~\citep{glegola2021mobilenet}.

Silva et al.~\citep{silva2017exploring} built a CNN based VAD model using audio spectrograms to detect speech in audio signals. Using the LeNet 5 CNN and the Half Total Error Rate metric, the proposed method outperformed several baseline VAD models in low, medium, and high noise conditions.  In an effort to further optimize VAD models in noisy conditions, ~\citep{lin2019optimizing} integrated a 2-layer bottleneck Denoising Autoencoder (DAE) with a CNN. The researchers carried out experiments using two different feature sets, MFCCs (Mel-Frequency Cepstral Coefficients) and filterbanks, and compared their performance in various Signal-to-Noise Ratio (SNR) conditions. The results demonstrated an improvement in classifying speech in high noise environments. In an attempt to measure human disturbance in ecological settings, ~\citep{cretois2022voice} proposed an alternative approach for acoustic VAD models. Researchers trained CNN models on synthetic datasets containing human voices mixed with typical background noises encountered in eco-acoustic data. By proposing a specialized preprocessing pipeline for audio augmentation, and synthetic dataset building, the results indicated the performance of a custom VGG11 model established a new state-of-the-art benchmark for VAD models in ecological settings. Despite the advances demonstrated in the aforementioned studies with respect to the accuracy of VAD models, the challenge of designing models that are suitable for real-time inference and deployment on edge devices remains a significant challenge.~\citep{alam2020lightweight} proposed a lightweight CNN with data augmentation and regularization techniques to improve the generalization ability of the model. Utilising the PreAct ResNet-18 architecture as a teacher, and log-scaled mel spectrogram as feature inputs, researchers trained a student model using response based distillation resulting in a lower equal error rate and latency from the distilled model. In a similar piece of research, ~\citep{dinkel2021voice} proposed a response based knowledge distillation  approach, where the teacher estimates the frame probability for each sound event and provided frame-level supervision to the student model, which was trained to then discriminate ground truth speech from non-speech labeled events. With the aim of deployment on embedded devices such as Raspberry Pi, the results indicated a 98\% reduction in parameters while outperforming the teacher model.

This study addresses the challenge of deploying deep learning models for ecological speech detection within the computational constraints of small, edge devices. These cost-effective and low-power devices struggle to efficiently run complex neural networks like EcoVAD, hampering real-time bioacoustic monitoring. To circumvent these challenges, our research focuses on applying knowledge distillation to create streamlined student models that parallel the larger EcoVAD teacher model’s performance. This approach is intended to overcome the inherent memory, latency, and computational limitations of such devices, while facilitating a more robust model capable of effective ecological monitoring.

\

\section{Methods}
\label{sec:sample:appendix}

In the current study, we build on the previous research discussed above which has proven instrumental in developing efficient, compact deep learning models suitable for deployment. We design and execute experiments to optimize deep neural networks for real-time speech detection. To achieve this objective, we investigate the suitability of MobileNetV3-Small-Pi~\citep{glegola2021mobilenet} model as a student architecture for EcoVAD~\citep{cretois2022voice}. The aforementioned studies also highlight the significance of specialized pre-processing, efficient lightweight architectures, and distillation techniques for optimizing VAD models for such deployment. Consequently, we employ different knowledge distillation techniques while incorporating variations of MobileNetV3-Small-Pi architecture to achieve optimal performance. Finally, we examine how reductions in parameters, floating-point operations per second (FLOPs), multiplications, and memory utilization in student VAD models influence the performance of the resulting architectures.

\subsection{Knowledge Distillation Techniques}

Hinton, Vinyals, and Dean~\citep{hinton2015distilling} first popularised the knowledge distillation method by training a smaller student network, using a teacher for distilled knowledge transfer. The method known as \textbf{response-based distillation}, trains the student to optimise the loss function based on the student and teachers softened outputs. While response based distillation allowed for “dark knowledge” to be distilled, depth is a critical aspect for feature representation learning~\citep{romero2014fitnets, gou2021knowledge}. 

In an attempt to distill intermediate representations, \citep{romero2014fitnets} introduced \textbf{feature-based distillation}, which trained a student network to optimise the loss function based on the students outputs and ground truth labels, along with the feature maps from an intermediary layer within the student and teacher, respectively. This method, which selects a teacher hidden layer as a “hint” layer, and student hidden layer as a “guide", improved the generalisation and accuracy of the student when compared to the teacher~\citep{gou2021knowledge}. 

While featured distillation allowed for deeper representation learning, the knowledge distilled is independent of outside data examples. Thus Park et al.~\citep{park2019relational}, introduced \textbf{relational knowledge distillation}, a method relying upon the relations between learned representations. This method trained the student network to optimise the loss function based on the angle wise, and distance wise relations between different data points, allowing the teacher to distill refined instance relations between layers and outputs of the model~\citep{park2019relational}.
\subsection{Model Architectures}
The teacher architecture used for knowledge distillation is based on a customised VGG11 architecture \citep{cretois2022voice}, adapted to process 128 x 128 single-color channel images, in contrast to the standard VGG11's handling of 224 x 224 RGB images. Significant modifications included the reconfiguration of input and output neurons, the introduction of batch normalization after each convolutional layer, and the implementation of a dropout strategy in fully connected layers to enhance the model's specificity for binary speech detection. All student architectures were based on MobileNetV3-Small-Pi (MSP) \citep{glegola2021mobilenet}. The student architectures maintained the differences implemented in \citep{glegola2021mobilenet}, with respect to MobileNetV3. More specifically, the adjustment from 5x5 filter with a 3x3 filter in the later convolution blocks, and adjustment from the hard-swish activation function to RELU. However, the architectural differences of the students differ from MSP in a number of ways. 

With the goal of analyzing the efficiency of student architectures, four different student designs were trained to measure the tradeoffs in accuracy and efficiency. The primary differences between these four architectures lie within the number of channels used in the convolutional and bottleneck layers, as well as the overall depth of the architecture. Student 1 starts with an initial 3x3 convolutional layer with 16 output channels, followed by a series of bottleneck layers with channels ranging from 16 to 512. Student 2 while similar to Student1, has a reduction in the number of bottleneck layers and a difference in the input channels prior to Adaptive Average pooling layer. The input channels are changed from 256 to 512 in this case. Student3 uses a smaller number of channels compared to aforementioned student architectures. The architecture, which starts with a a 3x3 convolutional layer, contains only 4 output channels in the initial bottleneck. The bottleneck layers in this architecture have channels ranging from 4 to 128, thus a reduction in capacity occurs as a result. The final student architecture, Student 4 maintains a similar structure to Student 3, with a 3x3 convolutional layer with 4 output channels in the initial bottleneck. However, the number of bottleneck layers is reduced in this case, leading to a more compact architecture with fewer layers. The channel sizes range from 4 to 64. 

Each student architecture maintains a similar final layer construction, which consists of an Adaptive Average Pooling layer, two 1x1 convolutional layers, and a flatten layer. The models also maintain the presence or absence of the SE block as in \citep{glegola2021mobilenet}. Furthermore, each student architecture maintains the same expansion ratio pattern, with the exception of Student 4. The differences in teacher and students architectures, which are highlighted in Table~\ref{tab:model-characteristics}, influence each respective model's capacity for feature extraction, and performance on the voice activity detection task.
\begin{sidewaystable}
\captionsetup{skip=5pt}  
\centering
\setlength{\tabcolsep}{11pt}  
\renewcommand{\arraystretch}{1.8}  
\begin{adjustbox}{center}  
\begin{tabular}{lccccccc}
\hline
Model     & Parameters & Layers & FLOPS       & Multiplications & Memory (MB) & Inference Time (s) \\ \hline
Teacher   & 59,568,769   & 20     & 2,485,390,000 & 1,242,700,000      & 227 & 0.17        \\
Student 1  & 4,662,017    & 215    & 388,459,000 & 194,230,000      & 17 & 0.038        \\
Student 2  & 2,930,177    & 179    & 337,257,000 & 168,628,000      & 11 & 0.042        \\
Student 3  & 502,793     & 179    & 27,353,400 & 13,676,700      & 1.91  & 0.0087        \\
Student 4  & 52,253      & 114    & 8,648,350 & 4,324,170      & 0.19  & 0.0050        \\ \hline
\end{tabular}
\end{adjustbox}
\caption{Summary of different teacher and student model characteristics.}
\label{tab:model-characteristics}
\end{sidewaystable}

\subsection{Dataset and Preprocessing}
The current study used three distinct datasets for the EcoVAD pre-processing pipeline: 

\textbf{The Soundscape-Dataset} \citep{cretois2022voice}, collected from the Bymarka forest near Trondheim, Norway, contains a total of 10 days of acoustic data recorded in files of 55 seconds at a sampling frequency of 44.1 kHz. From the initial 10 days of recordings, a subset of data was used for the EcoVAD preprocessing pipeline consisting of 9,037 raw audio signals from a continuous 5-day forest recording sampled with the same rate and intervals.

\textbf{The Libri-Speech Dataset} \citep{panayotov2015librispeech}, a corpus containing 1000 hours of 16kHz read English speech with a 1:1 male to female ratio was used for voice active detection. The data used for the EcoVAD pre-processing pipeline, was a subset from the corpus containing 360 hours, of which 200 hours of English reading speech with a 1:1 male/female ratio was extracted. 

\textbf{The Background Noise Dataset}, is a combination of ESC-50 dataset \citep{piczak2015esc} and BirdClef 2017 dataset \citep{kahl2017large}. The ESC-50 dataset, used for environmental sound classification, contains 2000 environmental recordings organized in 50 classes. For training, we subsetted the data to only include 1,600 recordings organized into 40 classes at 5-second intervals, removing human related sounds. The BirdClef 2017 Dataset, which includes audio recordings of various bird species, contains 36,496 audio recordings with 1500 species classes. Due to storage limitations, a subset of the dataset was used, accounting for 11,889 audio recordings belonging to 501 species. The three datasets, namely Soundscape, Libri-Speech, and Background Noise, were collectively utilized as inputs for the EcoVAD pre-processing pipeline.

The EcoVAD pre-processing pipeline \citep{cretois2022voice} was used to generate a synthetic data set consisting of 20,000 audio files, with a 1:1 distribution between speech and non speech audio files. The pipeline augments raw sound scape audio into processed 3-second sound scape audio clips, which are accompanied by ground truth labels denoting the presence or absence of speech. These processed 3-second sound scape audio clips, were augmented with speech, background, and bird species audio recordings to build an accurate representation of the ecological soundscape. To refine the raw audio signals into features for the speech detection task, the signals were converted into 128 x 128 Mel Spectrograms containing a single colour channel as in~\citep{cretois2022voice}. Mel-Spectrograms were then used as input into the student and teacher architectures for training.

\textbf{The Evaluation Playback Dataset} \citep{cretois2022voice}, is comprised of three-second audio clips representing the diverse environmental forest and grassland soundscapes. This dataset incorporates audio recordings of male, female, and child voices, both in speech and non-speech contexts, captured at distances of 1, 5, 10, and 20 meters. The playback dataset allows for the final evaluation and verification of the robustness of the various student models across distinct landscapes and at varying distances.

\subsection{Training and Evaluation}
The synthetic dataset generated for training each student model, was broken down into training, evaluation and test sets with the ratio of 60\%, 20\% and 20\% respectively. All models utilized in this study were subjected to a training process which involved a maximum of 50 epochs, employing batch sizes of 32. The number of inputs for each model was set to the Mel spectrograms feature dimensions, where the outputs for each model was set to one. Given the task is binary classification, this allows for the model to produce values between 0 and 1 in order to represent a prediction for speech detection. Furthermore, we use binary cross entropy with logits loss for the student losses, and binary cross entropy for the teacher loss function to accurately predict the binary classification task, and replicate the training procedure implemented in \citep{cretois2022voice}.\\ Moreover, after initial hyper-parameter testing, we found that the Adam optimiser \citep{kingma2014adam} was best suited for the optimization algorithm. Additionally, in each distillation experiment, we employed a learning rate of 0.001. The temperature parameter, which is used to soften the probability distribution of the logits, was set to 5. The alpha parameter, which controls the balance between the distillation loss and student loss in the total loss function, was set to 0.2. Finally, an early stopping method was used to prevent over-fitting. The method involved comparing the present validation loss with the best validation loss. Furthermore, a patience parameter was introduced and set to 3, in order to ensure that if the loss failed to improve over a predetermined number of epochs specified by the patience parameter, the training of the model would complete.

The evaluation metrics used to measure student model performance includes the F1 and AUC score. The F1 score, which can be seen as the harmonic mean of precision and recall, provides a single performance measurement for binary classifiers \citep{lipton2014optimal} On the other hand, the AUC score, which measures the area under a curve that represents the relationship between the true positive rate (TPR) and false positive rate (FPR), is also used for evaluation. Both the F1 score and AUC score were chosen to evaluate the student models based on the metrics employed in the training of the teacher model in \citep{cretois2022voice}.

\subsection{Software}
Python programming language (3.10.11) was used throughout the study. The pre-processing pipeline was developed using EcoVAD \citep{cretois2022voice}, which utilises Librosa v.0.8.1 \citep{mcfee2015librosa}, and Pydub v.0.25.1 \citep{robert2018pydub}, as the audio processing libraries. The data visualisations were done using matplotlib \citep{hunter2007matplotlib}. The pandas \citep{mckinney2010data}  and NumPy \citep{harris2020array} libraries were used for data loading and pre-processing. PyTorch (2.0.0) \citep{paszke2019pytorch} was used for developing the deep learning models. Scikit-learn \citep{pedregosa2011scikit} library was used for the evaluation of the models. The Google Colaboratory Environment \citep{bisong2019building} was used for training the models.

\section{Results}

\subsection{Refinement of Knowledge Distillation Techniques}

The performance of the student models reported using Average (Avg.) F1 and AUC scores varied across different distillation techniques. It is important to note that these averages were calculated over multiple runs without the use of fixed seeds, introducing a level of randomness into each experiment. Consequently, the average scores reflect a broad spectrum of performance under varying initial conditions, rather than a strictly controlled, consistent model behavior. Despite this, the results show a progression in model performance from soft target distillation to feature-based and ultimately to relational-based distillation (see Table~\ref{table:distillation_results}). For instance, in soft target distillation, the average AUC ranged between 0.9689 and 0.9831, while average F1 scores varied between 0.9372 and 0.9494.

\begin{table}[h!]
    \begin{center}
    \footnotesize
    \centering
    \setlength\tabcolsep{5pt} 
    \begin{tabular}{lccccccccccccccc}
    \toprule 
\bf Model		 & \multicolumn{2}{c}{\bf \makecell{Soft Target \\ Distillation}} & \multicolumn{2}{c}{\bf \makecell{Feature-Based \\ Distillation}} & \multicolumn{2}{c}{\bf \makecell{Relational-Based \\ Distillation}} \\
			 & Avg. AUC & Avg. F1 & Avg. AUC & Avg. F1  & Avg. AUC & Avg. F1  \\
\hline
Student 1            &  0.9831 & 0.9443 & 0.9888 & 0.9509 & 0.9896 & 0.9609 \\

Student 2      &  0.9824 &  0.9494 &  0.9893 &  0.9589 &  0.9898 &  0.9622 \\
Student 3      &  0.9810 &  0.9372 &  0.9870 &  0.9496 &  0.9856 &  0.9528 \\
Student 4      &  0.9689 &  0.9405 &  0.9883 &  0.9542 &  0.9880 &  0.9545 \\

    \bottomrule
    \end{tabular}
    \end{center}
    \caption{Table of results for different student models employing distinct distillation techniques. We report the Avg. AUC and Avg. F1 score of the student models, averaged over five runs, for each data set. \\}
    \label{table:distillation_results}
\end{table}

In the feature-based distillation experiment, an improvement was observed in both average AUC and F1 scores. AUC scores ranged from 0.9870 to 0.9893, and F1 scores varied between 0.9496 and 0.9589.The most significant improvement was seen in the relational-based distillation method, with average AUC scores ranging from 0.9856 to 0.9898 and F1 scores varying between 0.9528 and 0.9622. These results suggest that the refinement of knowledge distillation techniques can improve the performance of the resulting models.

\subsection{Impact of Parameter Reduction and Efficiency on Model Accuracy}
\begin{figure}[htbp]
  \centering
  \includegraphics[width=0.8\textwidth]{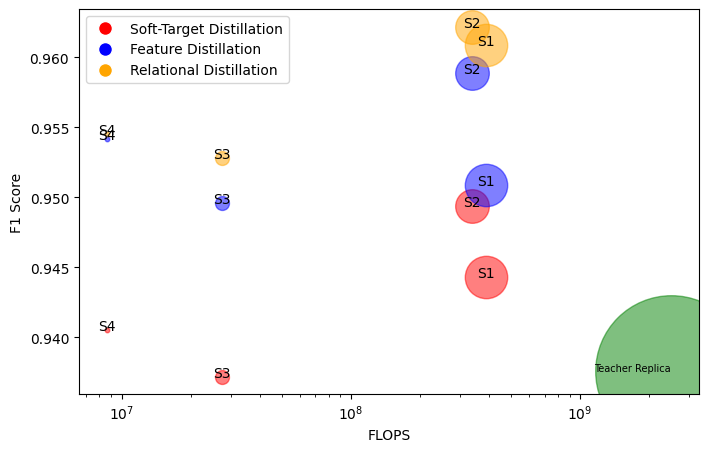}
  \caption{Varied distillation technique results per student with respect to FLOPS and Size. S1-S4 correspond to the four student models, while the Teacher Replica is the EcoVAD model. The size of circles corresponds to the number of parameters.}
  \label{fig:distillationviz}
\end{figure}

The reduction in parameters, FLOPs, multiplications, and memory utilization had varied accuracy across different distillation techniques (Figure \ref{fig:distillationviz}). Despite these reductions, the F1 scores of the student models did not significantly decrease when compared to the teacher-replica model. For instance, Student 1, with only 4,662,017 parameters and 388,459,000 FLOPs, achieved an Avg. F1 score of 0.9609 in the relational distillation method, which was higher than the teacher-replica model's F1 score of 0.9376.

The results indicate that the models are not in alignment with the assumption that a direct linear relationship exists between reductions in model parameters -- inclusive of floating-point
operations per second (FLOPs), multiplications, and memory utilization --
and model accuracy, as Student 2 and Student 4 outperform Student 1 and Student 3 respectively.

\subsection{Performance of Lightweight Student Models On Playback Dataset}

\begin{figure}[htbp]
  \centering
  \includegraphics[width=0.8\textwidth]{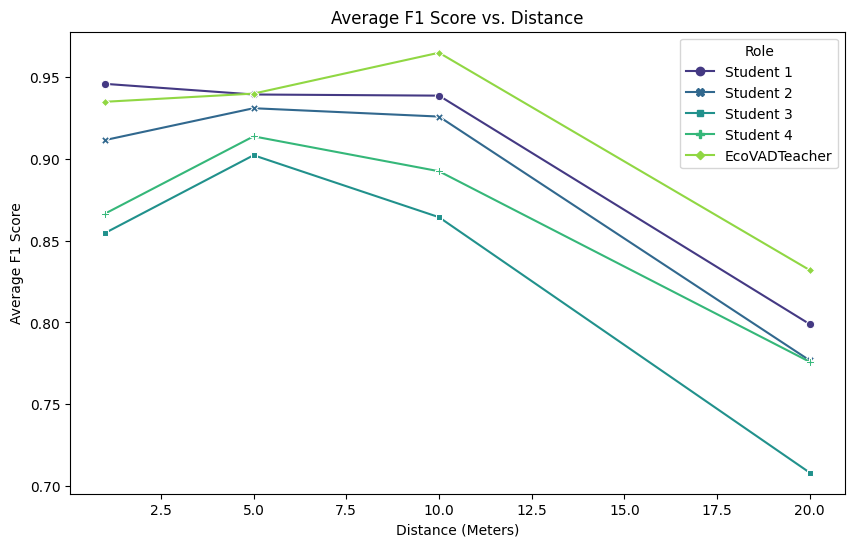}
  \caption{Avg. F1 scores based on distance on the playback evaluation data set for Relational based models}
  \label{fig:playbacks}
\end{figure}

In terms of performance, the student models demonstrated comparable, and in one instance, superior performance relative to the EcoVAD teacher model (Figure \ref{fig:playbacks}). For instance, Student 1 achieved an average F1 score of 0.94595, 0.93945, 0.93875, and 0.79895 at 1, 5, 10, and 20 meters, respectively, compared to the EcoVAD teacher model's average F1 scores of 0.93500, 0.94000, 0.96500, and 0.83200 at the same distances. 

Furthermore, while the Avg. F1 score across all distances for the EcoVAD model was .917, the student Avgs. were .905, .886, .832, and .862 for Students 1-4 respectively. These results indicate that efficient, lightweight student models can achieve comparable performance relative to the more complex EcoVAD teacher model.\\

These results highlight the potential of using knowledge distillation techniques for generating efficient, lightweight models for VAD tasks. Furthermore, these models maintain their accuracy despite significant reductions in parameters, FLOPs, multiplications, and memory utilization.

\section{Discussion}
The goal of the study was to build an efficient general purpose algorithm for voice detection in environmental audio by comparing different knowledge distillation techniques and student model architectures. More specifically, using the EcoVAD model \citep{cretois2022voice} as a teacher, and variations of MobileNetV3-Small-Pi \citep{glegola2021mobilenet} as student models, to compare knowledge distillation techniques for designing an efficient EcoVAD model. The results of this study were compared against the EcoVAD model on a playback dataset to evaluate the robustness of the efficient EcoVAD models on different landscapes with varying distances. 

The study demonstrates that efficient, lightweight student models can indeed achieve comparable performance relative to the EcoVAD teacher architecture using knowledge distillation and efficient student architectures. Student 1 and 2 maintained similar Avg. F1 scores on the playbacks dataset using the relational distillation models, when compared to the EcoVAD teacher model. This outcome supports the findings of previous research that distillation techniques can be used to create smaller, more efficient models without a significant reduction in accuracy \citep{gou2021knowledge}. Furthermore, the results from the different distillation techniques implemented have shown that a refinement in the knowledge distillation process increases student model performance. These results can be explained in the knowledge distillation literature \citep{gou2021knowledge}. Interestingly, in all distillation experiments Student 2 architecture outperformed Student 1 in the VAD task on the test dataset, however Student 1 outperformed Student 2 on the evaluation playback dataset. This could be the result of Student 2 being overfitted on the test set, however further testing would need to be done in order to determine if that is the case. 

The study also demonstrates that reductions in parameters, FLOPs, multiplications, and memory utilization do not necessarily result in a significant decrease in model accuracy. However, the results were not linear. The student models performance on the evaluation dataset demonstrated that while Student1 and Student2 outperformed the smaller models, Student4 consistently outperformed Student3 on both the test set, and playbacks dataset. This could be the result of certain architectural design features between student models, such as a difference in the expansion ratio and SE block implementation, however further testing would need to done in order to validate these claims. 
\section{Conclusions}
This study demonstrates that efficient student models can achieve comparable performance to EcoVAD. The findings indicate that MobileNetV
3-Small-Pi \citep{glegola2021mobilenet} can serve as a backbone for building efficient EcoVad models capable of achieving results comparable to EcoVAD teacher model \citep{cretois2022voice}. The study suggests that Student1 model has the potential to serve as a viable lightweight EcoVAD model for deployment on small devices for real time ecological monitoring.

The results of the current study are subject to certain limitations. The study incorporated a limited range of distillation techniques, therefore other distillation methods could serve to improve upon the current results. Additionally, the experiments ran were non-deterministic, therefore the implementation of a fixed seed could potentially enhance the reproducibility of these experiments. Moreover, portions of the data used to generate the synthetic dataset are proprietary, therefore restricted for research purposes only. Future research could explore the use of other distillation techniques, while further investigating different variations of the student EcoVAD models presented. Additionally, research could investigate the performance of these models in real-world settings.

The present study indicates that the optimization and design of efficient lightweight student models can lead results comparable to the larger EcoVAD model. While the current study is in no way a thorough investigation into efficient VAD model design, it can be considered a contribution towards the design of a efficient general purpose algorithm for voice detection in ecological settings. 
\appendix
 \bibliographystyle{elsarticle-num} 
 \bibliography{cas-refs}

\begin{thebibliography}{10}
\expandafter\ifx\csname url\endcsname\relax
  \def\url#1{\texttt{#1}}\fi
\expandafter\ifx\csname urlprefix\endcsname\relax\def\urlprefix{URL }\fi
\expandafter\ifx\csname href\endcsname\relax
  \def\href#1#2{#2} \def\path#1{#1}\fi

\bibitem{Stowell2022computational}
D.~Stowell, Computational bioacoustics with deep learning: a review and
  roadmap, PeerJ 10 (2022) e13152.

\bibitem{gong2021ast}
Y.~Gong, Y.-A. Chung, J.~Glass, Ast: Audio spectrogram transformer, arXiv
  preprint arXiv:2104.01778 (2021).

\bibitem{pan2022edgevits}
J.~Pan, A.~Bulat, F.~Tan, X.~Zhu, L.~Dudziak, H.~Li, G.~Tzimiropoulos,
  B.~Martinez, Edgevits: Competing light-weight cnns on mobile devices with
  vision transformers, in: Computer Vision--ECCV 2022: 17th European
  Conference, Tel Aviv, Israel, October 23--27, 2022, Proceedings, Part XI,
  Springer, 2022, pp. 294--311.
\newblock \href {https://doi.org/10.1007/978-3-031-03490-6_45}
  {\path{doi:10.1007/978-3-031-03490-6_45}}.

\bibitem{cretois2022voice}
B.~Cretois, C.~M. Rosten, S.~S. Sethi, Voice activity detection in eco-acoustic
  data enables privacy protection and is a proxy for human disturbance, Methods
  in Ecology and Evolution 13~(12) (2022) 2865--2874.

\bibitem{gaynor2018influence}
K.~M. Gaynor, C.~E. Hojnowski, N.~H. Carter, J.~S. Brashares, The influence of
  human disturbance on wildlife nocturnality, Science 360~(6394) (2018)
  1232--1235.

\bibitem{lewis2021human}
J.~S. Lewis, S.~Spaulding, H.~Swanson, W.~Keeley, A.~R. Gramza, S.~VandeWoude,
  K.~R. Crooks, Human activity influences wildlife populations and activity
  patterns: implications for spatial and temporal refuges, Ecosphere 12~(5)
  (2021) e03487.

\bibitem{hoke2021spatio}
K.~L. Hoke, N.~Hensley, J.~K. Kanwal, S.~Wasserman, N.~I. Morehouse,
  Spatio-temporal dynamics in animal communication: A special issue arising
  from a unique workshop-symposium model, Integrative and Comparative Biology
  61~(3) (2021) 783--786.

\bibitem{buxton2017noise}
R.~T. Buxton, M.~F. McKenna, D.~Mennitt, K.~Fristrup, K.~Crooks, L.~Angeloni,
  G.~Wittemyer, Noise pollution is pervasive in us protected areas, Science
  356~(6337) (2017) 531--533.

\bibitem{hill2019audiomoth}
A.~P. Hill, P.~Prince, J.~L. Snaddon, C.~P. Doncaster, A.~Rogers, Audiomoth: A
  low-cost acoustic device for monitoring biodiversity and the environment,
  HardwareX 6 (2019) e00073.

\bibitem{solomes2020efficient}
A.-M. Solomes, D.~Stowell, Efficient bird sound detection on the bela embedded
  system, in: ICASSP 2020-2020 IEEE International Conference on Acoustics,
  Speech and Signal Processing (ICASSP), IEEE, 2020, pp. 746--750.

\bibitem{gou2021knowledge}
J.~Gou, B.~Yu, S.~J. Maybank, D.~Tao, Knowledge distillation: A survey,
  International Journal of Computer Vision 129 (2021) 1789--1819.

\bibitem{hershey2017cnn}
S.~Hershey, S.~Chaudhuri, D.~P. Ellis, J.~F. Gemmeke, A.~Jansen, R.~C. Moore,
  M.~Plakal, D.~Platt, R.~A. Saurous, B.~Seybold, et~al., Cnn architectures for
  large-scale audio classification, in: 2017 ieee international conference on
  acoustics, speech and signal processing (ICASSP), IEEE, 2017, pp. 131--135.

\bibitem{howard2017mobilenets}
A.~G. Howard, M.~Zhu, B.~Chen, D.~Kalenichenko, W.~Wang, T.~Weyand,
  M.~Andreetto, H.~Adam, Mobilenets: Efficient convolutional neural networks
  for mobile vision applications, arXiv preprint arXiv:1704.04861 (2017).

\bibitem{sandler2018mobilenetv2}
M.~Sandler, A.~Howard, M.~Zhu, A.~Zhmoginov, L.-C. Chen, Mobilenetv2: Inverted
  residuals and linear bottlenecks, in: Proceedings of the IEEE conference on
  computer vision and pattern recognition, 2018, pp. 4510--4520.

\bibitem{hu2018squeeze}
J.~Hu, L.~Shen, G.~Sun, Squeeze-and-excitation networks, in: Proceedings of the
  IEEE conference on computer vision and pattern recognition, 2018, pp.
  7132--7141.

\bibitem{howard2019searching}
A.~Howard, M.~Sandler, G.~Chu, L.-C. Chen, B.~Chen, M.~Tan, W.~Wang, Y.~Zhu,
  R.~Pang, V.~Vasudevan, et~al., Searching for mobilenetv3, in: Proceedings of
  the IEEE/CVF international conference on computer vision, 2019, pp.
  1314--1324.

\bibitem{kang2022evaluation}
P.~Kang, A.~Somtham, An evaluation of modern accelerator-based edge devices for
  object detection applications, Mathematics 10~(22) (2022) 4299.

\bibitem{glegola2021mobilenet}
W.~Glego{\l}a, A.~Karpus, A.~Przyby{\l}ek, Mobilenet family tailored for
  raspberry pi, Procedia Computer Science 192 (2021) 2249--2258.

\bibitem{silva2017exploring}
D.~A. Silva, J.~A. Stuchi, R.~P.~V. Violato, L.~G.~D. Cuozzo, Exploring
  convolutional neural networks for voice activity detection, Cognitive
  technologies (2017) 37--47.

\bibitem{lin2019optimizing}
R.~Lin, C.~Costello, C.~Jankowski, V.~Mruthyunjaya, Optimizing voice activity
  detection for noisy conditions., in: INTERSPEECH, 2019, pp. 2030--2034.

\bibitem{alam2020lightweight}
T.~Alam, A.~Khan, Lightweight cnn for robust voice activity detection, in:
  Speech and Computer: 22nd International Conference, SPECOM 2020, St.
  Petersburg, Russia, October 7--9, 2020, Proceedings, Springer, 2020, pp.
  1--12.

\bibitem{dinkel2021voice}
H.~Dinkel, S.~Wang, X.~Xu, M.~Wu, K.~Yu, Voice activity detection in the wild:
  A data-driven approach using teacher-student training, IEEE/ACM Transactions
  on Audio, Speech, and Language Processing 29 (2021) 1542--1555.

\bibitem{hinton2015distilling}
G.~Hinton, O.~Vinyals, J.~Dean, Distilling the knowledge in a neural network,
  arXiv preprint arXiv:1503.02531 (2015).

\bibitem{romero2014fitnets}
A.~Romero, N.~Ballas, S.~E. Kahou, A.~Chassang, C.~Gatta, Y.~Bengio, Fitnets:
  Hints for thin deep nets, arXiv preprint arXiv:1412.6550 (2014).

\bibitem{park2019relational}
W.~Park, D.~Kim, Y.~Lu, M.~Cho, Relational knowledge distillation, in:
  Proceedings of the IEEE/CVF Conference on Computer Vision and Pattern
  Recognition, 2019, pp. 3967--3976.

\bibitem{panayotov2015librispeech}
V.~Panayotov, G.~Chen, D.~Povey, S.~Khudanpur, Librispeech: an asr corpus based
  on public domain audio books, in: 2015 IEEE international conference on
  acoustics, speech and signal processing (ICASSP), IEEE, 2015, pp. 5206--5210.

\bibitem{piczak2015esc}
K.~J. Piczak, Esc: Dataset for environmental sound classification, in:
  Proceedings of the 23rd ACM international conference on Multimedia, 2015, pp.
  1015--1018.

\bibitem{kahl2017large}
S.~Kahl, T.~Wilhelm-Stein, H.~Hussein, H.~Klinck, D.~Kowerko, M.~Ritter,
  M.~Eibl, Large-scale bird sound classification using convolutional neural
  networks., CLEF (working notes) 1866 (2017).

\bibitem{kingma2014adam}
D.~P. Kingma, J.~Ba, Adam: A method for stochastic optimization, arXiv preprint
  arXiv:1412.6980 (2014).

\bibitem{lipton2014optimal}
Z.~C. Lipton, C.~Elkan, B.~Naryanaswamy, Optimal thresholding of classifiers to
  maximize f1 measure, in: Machine Learning and Knowledge Discovery in
  Databases: European Conference, ECML PKDD 2014, Nancy, France, September
  15-19, 2014. Proceedings, Part II 14, Springer, 2014, pp. 225--239.

\bibitem{mcfee2015librosa}
B.~McFee, C.~Raffel, D.~Liang, D.~P. Ellis, M.~McVicar, E.~Battenberg,
  O.~Nieto, librosa: Audio and music signal analysis in python, in: Proceedings
  of the 14th python in science conference, Vol.~8, 2015, pp. 18--25.

\bibitem{robert2018pydub}
J.~Robert, Pydub: Manipulate audio with a simple and easy high level interface
  (2018).

\bibitem{hunter2007matplotlib}
J.~D. Hunter, Matplotlib: A 2d graphics environment, Computing in science \&
  engineering 9~(03) (2007) 90--95.

\bibitem{mckinney2010data}
W.~McKinney, et~al., Data structures for statistical computing in python, in:
  Proceedings of the 9th Python in Science Conference, Vol. 445, Austin, TX,
  2010, pp. 51--56.

\bibitem{harris2020array}
C.~R. Harris, K.~J. Millman, S.~J. Van Der~Walt, R.~Gommers, P.~Virtanen,
  D.~Cournapeau, E.~Wieser, J.~Taylor, S.~Berg, N.~J. Smith, et~al., Array
  programming with numpy, Nature 585~(7825) (2020) 357--362.

\bibitem{paszke2019pytorch}
A.~Paszke, S.~Gross, F.~Massa, A.~Lerer, J.~Bradbury, G.~Chanan, T.~Killeen,
  Z.~Lin, N.~Gimelshein, L.~Antiga, et~al., Pytorch: An imperative style,
  high-performance deep learning library, Advances in neural information
  processing systems 32 (2019).

\bibitem{pedregosa2011scikit}
F.~Pedregosa, G.~Varoquaux, A.~Gramfort, V.~Michel, B.~Thirion, O.~Grisel,
  M.~Blondel, P.~Prettenhofer, R.~Weiss, V.~Dubourg, et~al., Scikit-learn:
  Machine learning in python, the Journal of machine Learning research 12
  (2011) 2825--2830.

\bibitem{bisong2019building}
E.~Bisong, et~al., Building machine learning and deep learning models on Google
  cloud platform, Springer, 2019.

\end{thebibliography}





\end{document}